# High-order tensor neural network for iteration-free structure relaxation

Shaobo Yu[1], Haoting Zhang[1], Yu Han[1], Zhennan Zhang[1], Zhiyue Guo[1], Junjie Wang[1,*], Hao Gao[2,*], and Jian Sun[1,*]

[1] National Laboratory of Solid State Microstructures, School of Physics and Collaborative Innovation Center of Advanced Microstructures, Nanjing University, Nanjing 210093, China

[2] School of Physics, South China Normal University, Guangzhou 510006, China

## Abstract

Structure relaxation is important for the discovery of new materials, yet conventional *ab initio* optimization remains a major bottleneck in high-throughput screening workflows. Machine learning potentials have accelerated relaxation by orders of magnitude, but they still rely on iterative optimization and high-quality DFT force labels. Here, we present HotRelax, a high-order tensor message-passing neural network for one-shot, end-to-end prediction of relaxed structures. Trained directly on paired unrelaxed and relaxed structures, HotRelax requires no DFT force labels and predicts relaxed structures in a single forward pass, without iterative inference or post-processing. Across five diverse datasets spanning 3D bulk crystals, 2D layered materials and catalysts, HotRelax shows strong performance relative to state-of-the-art end-to-end relaxation models, achieving lower prediction errors on several benchmarks while maintaining a compact model size and efficient inference. Extensive DFT calculations further show that the predicted structures are close in energy to their DFT-relaxed counterparts. When integrated into catalytic workflows, HotRelax also improves the accuracy and generalization of relaxed-state energy prediction models. Together, these results support HotRelax as an efficient and widely applicable framework for end-to-end structure relaxation, with strong potential to accelerate high-throughput materials discovery.

---

* Corresponding author: J.S. (jiansun@nju.edu.cn); H. G. (gaohao@m.scnu.edu.cn); J. W. (wangjunjie@nju.edu.cn)

# Introduction

Structure relaxation is fundamental to the discovery and development of novel functional materials. In workflows that explore vast chemical spaces, whether through high-throughput screening[1,2] or global-search strategies such as genetic algorithms[3–8] and particle swarm optimization[9,10], relaxing the initial structure is a central step. The goal is to identify stable or metastable configurations corresponding to local minima on the potential energy surface (PES). For decades, density functional theory (DFT) has been the standard approach for this task. By self-consistently solving the Kohn-Sham equations, DFT accurately describes interatomic interactions and guides iterative relaxation toward energy minima via gradient-based algorithms[11,12]. Despite its accuracy and wide applicability, however, DFT-based structure optimization remains a major bottleneck in large-scale studies. A standard DFT relaxation involves two nested iterative procedures. In the inner loop, self-consistent field (SCF) calculations solve the Kohn-Sham equations for a fixed atomic configuration to obtain the electronic ground state, total energy, and atomic forces. In the outer loop, these forces are used to update atomic positions and lattice parameters until the structure converges to a local minimum on the PES. Because each geometry update in the outer loop requires repeated SCF calculations in the inner loop, the overall computational cost rises rapidly with both system size and the number of relaxation steps. Conventional DFT therefore becomes impractical for high-throughput screening campaigns involving tens of thousands to millions of candidate structures[13].

In recent years, machine learning (ML) methods, particularly machine learning potentials (MLPs), have become useful tools for reducing this DFT bottleneck[14]. Trained on DFT-derived energies and forces, MLPs deliver near-first-principles accuracy at far lower computational cost, replacing iterative SCF calculations by one-step prediction, thus accelerating structure relaxation and molecular dynamics by orders of magnitude[15–25]. Yet MLPs remain surrogate models of PES[26], rather than direct mappers from initial structures to relaxed ones. Even when they are trained extremely well, relaxation at inference time still requires iterative gradient-based optimization, which limits efficiency and prevents one-shot prediction. A second limitation is data availability. MLPs require high-quality datasets with dense DFT energy and force labels, whereas most public databases provide only final relaxed configurations and do not include intermediate trajectory and force data[27]. This mismatch severely restricts MLPs training and generalization across diverse material families and complex systems.

To overcome the limitations of iterative MLPs-based approaches, a new class of ML frameworks aims to predict relaxed structures in a non-iterative manner. The central idea is to learn an end-to-end mapping from an unrelaxed structure to its relaxed counterpart, thereby bypassing iterative geometry optimization altogether. Early studies showed that this scheme can work. For example, Yoon et al.[28] proposed DOGSS, a differentiable optimization framework that learns optimal harmonic force-field parameters to guide initial configurations toward near relaxed-state structures. Kim et al. introduced Cryslator[29], which learns a mapping between the latent feature spaces of unrelaxed and relaxed structures and thereby infers the relaxed geometry indirectly. Building on these foundations, Yang et al. proposed DeepRelax[27] and $E^3$Relax[30]. DeepRelax reconstructs the relaxed structure by predicting interatomic distances and atomic displacements from the unrelaxed configuration[27]. However, its raw outputs still require post-hoc geometric optimization to recover the final atomic coordinates, limiting its end-to-end capability. By contrast, $E^3$Relax, currently among the strongest end-to-end frameworks[30], progressively updates atomic coordinates and lattice vectors through layer-wise updates within a single forward pass and achieves state-of-the-art (SOTA) performance across several tasks. This design, however, increases architectural complexity and yields a larger model. To model lattice deformation explicitly, $E^3$Relax also introduces three dedicated lattice nodes that interact with all atom nodes. Although effective in practice, this design is less physically transparent because it couples quantities with different geometric meanings. This observation motivates the search for alternative ways to incorporate lattice information into end-to-end relaxation models.

Here we present HotRelax, a high-order tensor message passing neural network for one-shot, end-to-end prediction of relaxed crystal structures. HotRelax addresses several key limitations of existing methods: it requires no DFT force labels, learning directly from paired unrelaxed and relaxed structures; it enables single-step prediction without iterative inference or post-processing; and it avoids complex architectural designs such as global lattice nodes while retaining high accuracy. We benchmark HotRelax on five diverse datasets: the X-Mn-O (X = Mg, Ca, Ba and Sr) dataset[29,31], the Materials Project (MP) dataset[32], the Computational 2D Materials Database (C2DB)[33–35], the 2D material defects[36,37] (2DMD) and the Open Catalyst 2020 (OC20) dataset[38]. Across these benchmarks, HotRelax shows strong structure-prediction performance while maintaining a compact model size and inference speed comparable to $E^3$Relax. DFT validation further shows that its predicted structures are closer in energy to DFT-relaxed references. When integrated into the OC20 workflow[38], HotRelax serves as a structure preprocessing module that generates predicted relaxed configurations; energy prediction models trained on these configurations outperform those trained directly on unrelaxed structures, improving downstream catalyst screening pipeline. These results suggest the potential of HotRelax for high-throughput screening.

# Results

## Decoupling of coordinates and lattice

The choice of learning objective directly influences model convergence and prediction accuracy. In an early direct-relaxation study, Kim et al. explored how to predict structural changes between initial and relaxed crystals[29]. Rather than learning the full relaxed structure directly from the initial one, Cryslator adopts an indirect latent-space translation strategy[29]. Specifically, it first pre-trains a graph encoder and regressor on relaxed structures using formation energy as the target, thereby learning a latent representation of the relaxed structural domain. It then learns a translation from the latent features of unrelaxed structures to those of relaxed structures and finally uses the translated features together with the initial structure to predict the relaxed geometry. This design avoids learning the full structural mapping entirely from scratch, but it remains an indirect route to relaxed-structure prediction.

Subsequent work has focused on target representations that preserve direct relaxation prediction while improving training stability. A representative example is DeepRelax[27], which reformulates structure relaxation as the prediction of several geometrically meaningful quantities, including atomic displacements, relaxed interatomic distances, and the relaxed lattice matrix, rather than directly regressing only the final Cartesian coordinates. This reformulation retains key geometric information and provides a more structured learning objective for reconstructing the relaxed configuration. By contrast, $E^3$Relax directly predicts atomic coordinates and lattice parameters but uses a residual learning mechanism[30], in which each GNN layer predicts only the residual correction from the previous layer's output relative to the target relaxed structure. Although this progressive approximation eases high-dimensional regression, it still relies on progressive residual updates within the network.

The main difficulty of direct structure relaxation therefore lies not only in the high dimensionality of the output space. A more fundamental challenge is that atomic rearrangement and lattice deformation are entangled in Cartesian coordinates. For a crystal structure, the relaxed Cartesian coordinates can be written as

$$\boldsymbol{R}^r = F^r \cdot \boldsymbol{L}^r \quad (1)$$

where $F^r$ denotes the relaxed fractional coordinates and $\boldsymbol{L}^r$ denotes the relaxed lattice matrix. This expression shows that the final Cartesian coordinates are jointly determined by two distinct factors, namely the intra-cell atomic arrangement encoded by $F^r$ and the global cell geometry encoded by $\boldsymbol{L}^r$. Directly regressing $\boldsymbol{R}^r$ in Cartesian space therefore forces the model to learn these two effects simultaneously.

Motivated by the natural separation between fractional coordinates and lattice vectors in crystal representations, we introduce a decoupling of coordinates and lattice (DCL) strategy as a reparameterization of the relaxation target. Instead of directly predicting the relaxed Cartesian coordinates under the relaxed lattice, we first define an auxiliary coordinate target under the initial lattice

$$\boldsymbol{R}_{DCL}^r = F^r \cdot \boldsymbol{L}^u \tag{2}$$

where $\boldsymbol{L}^u$ is the lattice matrix of the unrelaxed structure. This representation preserves the relaxed atomic arrangement while removing the contribution of lattice deformation from the coordinate target. The learning objective is then decomposed as

$$\Delta\boldsymbol{R} = \boldsymbol{R}_{DCL}^r - \boldsymbol{R}^u = (F^r - F^u) \cdot \boldsymbol{L}^u \tag{3}$$

$$\Delta\boldsymbol{L} = \boldsymbol{L}^r - \boldsymbol{L}^u \tag{4}$$

Here $\Delta\boldsymbol{R}$ describes atomic relaxation within a fixed reference cell, whereas $\Delta\boldsymbol{L}$ captures lattice deformation. The original coupled prediction problem is thus rewritten as two more structured subproblems.

This reparameterization does not alter the final physical target. After prediction, the relaxed structure is reconstructed through

$$\hat{F}^r = \left(\boldsymbol{R}^u + \Delta\widehat{\boldsymbol{R}}\right) \cdot (\boldsymbol{L}^u)^{-1} \tag{5}$$

$$\widehat{\boldsymbol{L}}^r = \boldsymbol{L}^u + \Delta\widehat{\boldsymbol{L}} \tag{6}$$

$$\widehat{\boldsymbol{R}}^r = \hat{F}^r \cdot \widehat{\boldsymbol{L}}^r \tag{7}$$

DCL can therefore be understood as a physically motivated reparameterization that better reflects the intrinsic geometry of crystals, in which fractional coordinates describe intra-cell atomic arrangement and lattice vectors describe periodic cell geometry.

To isolate the effect of DCL from other architectural factors, we evaluated it using PaiNN[20] as a simple and controlled equivariant backbone. This comparison is also relevant to DeepRelax[27]. The core architecture of DeepRelax[27], PaEGNN, follows the same scalar-vector equivariant message-passing scheme as PaiNN[20] while adding periodic boundary encoding and additional geometric targets for subsequent EDG-based structure reconstruction. Comparing PaiNN[20] with PaiNN-DCL therefore isolates the contribution of DCL itself, whereas comparison with DeepRelax[27] indicates whether this decoupled learning objective can already achieve competitive performance relative to a more elaborate PaiNN-based relaxation framework.

As shown in Table 1, the vanilla PaiNN[20] exhibits substantially larger coordinate prediction error. After introducing DCL, the same backbone attains markedly improved accuracy and becomes competitive with DeepRelax[27]. These results suggest that the benefit does not arise solely from more elaborate geometric targets or reconstruction procedures, but already from rewriting the learning objective into a better-conditioned form. We therefore adopt the targets defined in Eqs. (3) and (4) for coordinate and lattice prediction in HotRelax.

## The architecture of HotRelax

As an end-to-end deep learning framework, HotRelax takes the element types, the coordinate matrix, and the lattice vectors of initial structures as inputs, and directly outputs atomic displacements and lattice adjustments that reconstruct the relaxed structure in a single step. Unlike existing methods that rely on iterative optimization or multi-layer residual approximations, HotRelax is designed to perform one-shot structural relaxation.

Figure 1 presents the architecture of HotRelax, which builds on the High-order Tensor message Passing interatomic Potential (HotPP)[24]. Here we adopt HotPP as the equivariant backbone and adapt it for direct prediction of relaxed structures. Scalar, vector, and second-order tensor features encode complementary geometric information in crystals, and these equivariant representations are refined through stacked message-passing layers. A task-specific readout layer then decodes atomic displacements and lattice vector variations for one-shot relaxation prediction.

At the architectural level, the key distinction between HotRelax, the original HotPP[24] model, and earlier end-to-end relaxation frameworks lies in how lattice information is incorporated throughout the network. Predicting global quantities such as lattice vectors is a central challenge in end-to-end structure relaxation, because lattice deformation reflects a collective structural response rather than a purely local atomic displacement. Existing models address this difficulty in different ways. In DeepRelax[27], lattice information is introduced in the output branch for relaxed-lattice prediction by combining encoded input-lattice features with pooled atomic representations. In $E^3$Relax[30], lattice deformation is modeled by introducing three explicit lattice nodes that interact with atom nodes throughout the network. HotRelax also maintains layer-wise lattice awareness, but in a different manner. Rather than representing lattice vectors as separate nodes, it embeds lattice vectors directly into the geometric operations of message passing across all propagation layers. In this way, global lattice information continuously informs interatomic feature refinement without requiring dedicated lattice nodes, reducing architectural overhead while preserving the capacity to model long-range periodic effects.

HotRelax is also designed to preserve equivariance throughout the network. Nonlinear operations are applied only to rotationally invariant scalar (0th-order) features, whereas updates for equivariant vectors (1st-order) and tensors (2nd-order) are performed through linear transformations and invariant scalar-modulated weighting. This design ensures that all features transform correctly under arbitrary Euclidean transformations. Moreover, atomic displacements and lattice variations are decoded solely from the equivariant vector outputs of the final interaction layer, providing end-to-end physical self-consistency across coordinate systems.

## The performance of HotRelax

To evaluate HotRelax comprehensively for end-to-end structural relaxation, we designed systematic benchmarks spanning representative 3D bulk and 2D layered materials. For 3D systems, we used two complementary datasets: the X-Mn-O oxide system[29,31], which provides a stringent test of metal oxides with complex relaxation behaviors, and the MP dataset[32], a broad benchmark of inorganic crystals spanning diverse compositions and symmetries. For 2D materials, we used the C2DB[33–35] dataset, which contains DFT-validated structures of diverse monolayers, and the 2DMD[36,37] dataset, which contains point-defect structures of several widely studied 2D materials.

Across these four benchmark datasets, we compared HotRelax with representative end-to-end relaxation frameworks, including DeepRelax[27] and $E^3$Relax[30] for the first three datasets, and E3Relax[30] and DefiNet[39] for 2DMD[36,37], together with a naive dummy baseline for reference. To ensure fairness and reproducibility, all competing models were evaluated using the same data splitting strategy and similar evaluation metrics. Notably, HotRelax was trained with an identical set of hyperparameters across all datasets, without any dataset-specific tuning or system-specific optimization. For reference, the numbers of trainable parameters were fixed across benchmarks at 15.8 M for DeepRelax[27], 74.6 M for $E^3$Relax[30], and 11.6 M for HotRelax.

## 3D crystal structure dataset

We first validated HotRelax on the X-Mn-O oxide dataset, a standard benchmark for end-to-end crystal relaxation that has been widely used in prior studies[27,29–31,40,41] and therefore enables direct performance comparison. Curated from MP, this dataset comprises ternary oxides of Mn, O, and alkaline-earth metals (X = Mg, Ca, Sr, Ba). The dataset is challenging because the initial configurations can deviate substantially from the relaxed structures (detailed in Supplementary Note 1), placing many geometries far from the local potential energy minimum. Such large displacements require models to learn intrinsic relaxation behavior rather than merely applying minor corrections, making this dataset a rigorous test of generalization and extrapolation.

The X-Mn-O dataset consists of 28,579 paired unrelaxed and DFT-relaxed structures. To ensure direct and fair comparison with prior models, we strictly followed the same data splitting protocol used in those studies[27,29,30]. Specifically, the dataset was randomly partitioned into training (22,863 pairs), validation (2,858 pairs), and test (2,858 pairs) sets according to an 8:1:1 ratio, guaranteeing that each subset exactly matches the configurations employed in prior studies. For quantitative evaluation, we adopted three metrics originally defined in DeepRelax[27]: mean absolute error (MAE) of Cartesian coordinates, MAE of lattice vectors, and structure match rate (MR) calculated by Pymatgen[42] with default parameters (ltol = 0.2, stol = 0.3).

Table 2 presents the performance comparison on the X-Mn-O test set. Details of the training for HotRelax are provided in Supplementary Note 2. The results demonstrate that HotRelax achieves strong performance in prediction accuracy. Relative to the previous SOTA model, HotRelax achieves an approximately 12% reduction in lattice MAE and an approximately 8% reduction in atomic coordinate MAE. Importantly, these gains do not come at the expense of efficiency; HotRelax uses fewer trainable parameters while maintaining comparable inference time.

To further assess the generalization and broad applicability, we next benchmarked HotRelax on the MP dataset[32], a standard benchmark widely used for inorganic crystal structure modeling. This dataset comprises 62,724 high-quality unrelaxed and relaxed structure pairs, spanning a diverse range of elements, chemical compositions, crystal systems, and space groups. To ensure direct comparability with prior models, we strictly followed the established data splitting protocol, partitioning the MP dataset into training (56,452 pairs), validation (3,136 pairs), and test (3,136 pairs) sets using an 18:1:1 ratio, identical to the splits used in previous benchmark studies[27,29,30]. In this dataset, the structural deviations between unrelaxed and relaxed configurations are considerably smaller than those in X-Mn-O (detailed in Supplementary Note 1), with most initial geometries already near local minima on the PES. This characteristic inherently limits the headroom for performance improvement and makes further reductions in prediction error more challenging to achieve.

Nevertheless, the results in Table 2 show that HotRelax still achieves strong performance on this benchmark. Relative to the previous SOTA model, it delivers an approximately 5% reduction in both atomic coordinate MAE and lattice vector MAE, yielding improvements in structure prediction accuracy. These results suggest that HotRelax captures the physical and chemical principles governing structural relaxation across chemically diverse crystalline systems.

## 2D crystal structure dataset

We next conducted systematic benchmarking on the two-dimensional material database C2DB[33–35]. As one of the most comprehensive and widely adopted benchmarks for computational 2D materials research, C2DB comprises 11,581 high-quality unrelaxed and relaxed structure pairs, spanning 62 chemical elements from light to heavy elements and including diverse 2D configurations such as hexagonal honeycomb lattices, square lattices, and transition metal dichalcogenide structures.

To ensure direct comparability with prior models, we strictly followed the established data splitting protocol[27,30]. The dataset was randomly partitioned into training (6,948 pairs), validation (2,317 pairs), and test (2,316 pairs) sets using a fixed 6:2:2 ratio, ensuring that the structure configurations and chemical distributions exactly match those used in benchmark studies[27,30].

Table 3 presents a comparison of HotRelax against competing models on the C2DB test set. The results clearly show that HotRelax achieves strong performance in prediction accuracy. Specifically, relative to the previous SOTA model, HotRelax achieves an approximately 4% reduction in coordinate MAE and an approximately 5% reduction in lattice MAE.

To further evaluate the performance of HotRelax on 2D materials, we benchmarked it on the 2DMD dataset[36,37]. The 2DMD dataset contains defect properties for a set of representative 2D materials, including $MoS_2$, $WSe_2$, hBN, GaSe, InSe, and black phosphorus (BP). With respect to defect density, 2DMD consists of two parts: one comprises low-density, structured defect configurations, with 5,933 structures each for $MoS_2$ and $WSe_2$; the other comprises high-density, randomly configured defects, with 500 structures per material, yielding a total of 3,000 structures across all six materials.

Because the low-density defect unrelaxed structures closely resemble their relaxed counterparts, differing primarily near the defect sites, we chose to benchmark directly on the more challenging high-density defect configurations. The dataset was randomly partitioned into training (2,400 pairs), validation (300 pairs), and test (300 pairs) sets using a fixed 8:1:1 ratio, identical to that employed in the previous study[39].

For quantitative evaluation, we adopted metrics originally defined in DefiNet[39]: coordinate MAE between unrelaxed and relaxed structures, and the localized MAE statistics near the defect sites. Specifically, atoms within a radius of $x$ Å from the defect sites are denoted as $A_x$, where $x$ takes values of 3, 4, 5, and 6. For example, the coordinate MAE for $A_5$ is calculated using only atoms that lie within a 5 Å radius of the defect site.

Figure 2 presents the results of HotRelax and competing models on the high-density defect structures from the 2DMD dataset. For clarity, we show results for the two materials that exhibit the largest differences between unrelaxed and relaxed structures; the complete results for all six materials are provided in Supplementary Note 4. $E^3$Relax[30], DefiNet[39], and HotRelax all achieve markedly improved performance over the Dummy model. Notably, HotRelax performs nearly on par with DefiNet, which is the current SOTA single-step ML model for defect structures[39]. This comparison is particularly informative because DefiNet is specifically designed for defect structures, and its strategies for processing initial structures and encoding defect sites are not readily transferable to generic crystal structures. By contrast, HotRelax attains comparable performance without any defect-specific processing, indicating stronger generalization capability.

## DFT validations

Total energy is a direct first-principles measure of structure stability. To test whether the structures predicted by HotRelax are also reasonable in energy, we performed DFT single-point energy calculations using VASP[43–46] on the X-Mn-O and MP datasets. For each dataset, we compared the energies of the initial unrelaxed structures, the DFT-relaxed structures, and the HotRelax-predicted structures without further optimization.

For the X-Mn-O dataset, we focused on 1,007 challenging test cases whose initial structures do not match the DFT-relaxed structures. As shown in Figure 3a, the HotRelax-predicted structures are much closer in energy to the DFT-relaxed references than the initial structures. The average energy above the relaxed state decreases from 32.49 eV for the initial structures to 2.95 eV for the HotRelax-predicted structures. For the MP dataset, Figure 3b shows that the HotRelax-predicted structures also have lower energies than the initial structures across the test set. These results support the physical reasonableness of the predicted structures. We also carried out an ablation study, and the details are provided in Supplementary Note 5.

## Apply HotRelax to catalysts

Discovering and optimizing catalysts are essential for advancing renewable energy processes and storage technologies[47]. However, the high computational cost of DFT-based geometric optimization for large-scale catalysts severely limits catalyst screening efficiency and novel catalyst development[48,49]. To overcome this bottleneck, we evaluate HotRelax on the Open Catalyst 2020 (OC20)[38] dataset, a comprehensive and widely adopted benchmark in computational catalysis. HotRelax trained on OC20 generates one-shot structure predictions that can be used to train ML models for fast and accurate prediction of DFT-relaxed energies, reducing the need for costly DFT relaxations during screening.

### Benchmark

We adopt the Initial Structure to Relaxed Energy (IS2RE)[38] task as our benchmark. IS2RE is a standard task in computational catalysis, designed to assess a model's ability to predict the energy of a DFT-relaxed structure directly from its initial, unrelaxed configuration. This task is particularly challenging because catalyst interfaces are structurally and chemically complex. The model should predict the energy of the DFT-relaxed state from the initial adsorbate-substrate geometry, without access to relaxation trajectories. Because the mapping from an initial configuration to its relaxed-state energy is highly nontrivial on a high dimensional PES, IS2RE remains more difficult than bulk-structure prediction tasks.

The IS2RE task provides a training set of 460,328 paired initial and relaxed structures with corresponding relaxed-state energies[38]. As a comprehensive benchmark, the dataset covers numerous materials, adsorbates, and surfaces. For evaluation, the IS2RE dataset defines four validation and test splits with increasing generalization difficulty[38]: in-domain (ID), out-of-domain adsorbates (OOD-ads), out-of-domain catalysts (OOD-cat), and both unseen adsorbates and unseen catalysts (OOD-both). Each split contains approximately 25k samples, leading to a total of approximately 100k samples for the full validation set and approximately 100k samples for the full test set, respectively.

In standard catalyst relaxation workflows, only the surface and adsorbate atoms undergo substantial structural changes, whereas most substrate atoms remain fixed at bulk positions. To exploit this characteristic, we introduced targeted adaptations in both training and evaluation pipelines. Specifically, for model training, we incorporated label embeddings into the scalar feature initialization stage. Atoms undergoing significant position changes are labeled "1", whereas fixed atoms are labeled "0". This design directs the model's learning focus toward active-site atoms, reducing emphasis on fixed bulk atoms that undergo little or no structural change during relaxation, which can improve training efficiency. All other hyperparameters and network architectures were kept strictly identical to those used for previous crystal structure datasets. For model evaluation, we introduce the active-site atomic coordinate MAE (AMAE) as a core evaluation metric, which computes prediction errors exclusively for the surface and adsorbate atoms. This enables precise quantification of model performance in the active region that governs catalytic behavior and the relaxed-state energy. We also report the structure MR for consistency with prior bulk-crystal experiments.

For benchmarking, we trained $E^3$Relax as the competing baseline. All models were trained and validated on the identical OC20 splits using the same evaluation pipeline, so that the observed performance differences primarily reflect model design rather than data partitioning. Note that the relaxed structures for the official test set are not public, thus we use the values across validation splits for benchmarking.

Table 4 presents the comparison of HotRelax against competing models on the four validation subsets. HotRelax achieves strong performance in prediction accuracy, indicating consistent generalization across different catalyst and adsorbate domains. Compared with the Dummy baseline, HotRelax reduces coordinate AMAE by 39.9% and improves structure MR by 118.9% on average, further highlighting its effectiveness.

As an example, Figure 4 visually demonstrates HotRelax's strong performance in predicting DFT-relaxed catalyst structures. Marked structural differences are evident between the unrelaxed structure and the DFT-relaxed structure. Relative to the relaxed reference, the unrelaxed structure yields a normalized root-mean-square displacement (RMSD) of 0.0525 and a maximum atomic displacement (MaxD) of 0.36 Å, as calculated using Pymatgen[42]. For comparison, geometry optimization of the unrelaxed structure using MLPs (here implemented with EquiformerV2 (31M)[23]) requires 78 iterations to converge to a configuration matching the relaxed reference, with a total runtime of ~5 seconds on a single GPU. By contrast, HotRelax predicts a configuration in agreement with the relaxed reference in a single step, with a runtime of only 0.02 s on a single GPU. These results highlight the potential of HotRelax for accelerating the prediction of stable catalysts and their corresponding energies.

## Improving energy prediction

We next applied HotRelax as an explicit structure preprocessing module to enable accurate and efficient energy prediction for catalysts. In the original IS2RE framework (Figure 5a), a GNN takes the unrelaxed configuration as input and aims to learn an implicit mapping from this initial structure to the energy of the DFT-relaxed structure. However, this framework suffers from a fundamental physical limitation. The mapping between the initial unrelaxed structure and the target relaxed-state energy is nontrivial because the energy label is associated with the DFT-relaxed configuration rather than the initial geometry itself. Without explicit access to relaxed structures, the GNN must infer this relationship implicitly, which can limit prediction accuracy and generalization.

To address this issue, we propose a two-stage prediction framework based on HotRelax (Figure 5b). This framework provides the downstream energy prediction model with predicted configurations that closely approximate the DFT-relaxed structures, thereby providing a more direct basis for the PS2RE (predicted structure to relaxed energy) mapping and reducing the learning difficulty of the energy prediction task.

The workflow proceeds as follows. First, we pre-train a HotRelax model on the OC20 training set using paired unrelaxed-relaxed structures. This pre-trained model enables one-shot end-to-end prediction of the relaxed configuration directly from an unrelaxed initial structure without iterative optimization. Second, we feed all unrelaxed initial structures from the training set into the pre-trained HotRelax model to generate corresponding predicted configurations. These predicted structures, paired with the relaxed-state energies, form a physically consistent training dataset for the GNN-based energy prediction model. Critically, the energy prediction model is trained exclusively on HotRelax-predicted structures rather than on the original unrelaxed initial structures. Third, for an unseen unrelaxed initial structure, we first use the pre-trained HotRelax model to obtain its predicted configuration and then feed this structure into the trained energy prediction model to produce the final relaxed-state energy prediction. Although this two-stage framework introduces an additional inference step compared with the conventional IS2RE, the added computational cost is negligible, as validated in our previous experiments. In return, the framework could obtain clear gains in relaxed-state energy prediction accuracy.

To validate the performance of our proposed framework, we conducted a systematic comparison experiment. We selected two GNN architectures as backbones for the energy prediction model, one is PaiNN[20], and the other is EquiformerV2[23], allowing us to assess whether the observed trend is consistent across different GNN backbones. To ensure fairness and reliable comparison, we implemented strict single-variable control. For each selected GNN backbone, we trained two parallel variants under identical settings, with the only difference being the input structures. The baseline variant followed the conventional IS2RE paradigm, using the original unrelaxed structures as input, whereas the HotRelax-augmented variant adopted our proposed two-stage framework, using the configuration predicted by HotRelax as input. Both variants of the same GNN backbone used identical training protocols, optimization hyperparameters, and evaluation pipelines. Details of the training are provided in Supplementary Note 7.

Table 5 presents a comparison of all models across the four validation subsets and the average values across four test subsets (obtained from Hugging Face[52]) of the IS2RE task. The results show that our proposed HotRelax-augmented prediction framework outperforms the conventional IS2RE paradigm across all scenarios and for both tested GNN backbones. This finding supports the use of HotRelax as a structural preprocessing module for catalyst energy prediction tasks. The performance improvement also follows a physically meaningful trend across subsets of increasing generalization difficulty. Using the EquiformerV2[23] backbone as an example, the two-stage framework reduces energy MAE by approximately 11% in the ID subset. In contrast, the improvement becomes larger in the more challenging OOD subsets, reaching approximately 17% in OOD-ads and approximately 20% in OOD-both, where the model encounters completely unseen adsorbate species or catalyst substrates. This trend suggests that, when an energy prediction model is applied to an unfamiliar chemical space, its predictive performance becomes more sensitive to quality of the input structures. In the conventional IS2RE paradigm, the model must learn a highly indirect mapping from an OOD unrelaxed structure to the relaxed-state energy, which limits its generalization capability. By contrast, HotRelax transforms the unrelaxed structure into a configuration that closely approximates the relaxed structures, thereby reducing the distribution shift between the training data and inference inputs. This physically motivated preprocessing is consistent with the observed improvement in generalization to unseen chemical systems.

## Discussion

In this work, we developed HotRelax, a high-order tensor message passing neural network that enables one-shot, end-to-end prediction of relaxed crystal structures. Across diverse material systems, including 3D bulk metal oxides, general inorganic crystals, 2D layered materials and catalyst interfaces, HotRelax achieves strong performance without dataset-specific hyperparameter tuning. This performance is likely supported by three physically informed design choices. First, the DCL strategy decouples the optimization of atomic coordinates and lattice vectors, allowing the model to learn local atomic displacements and global lattice deformation more clearly and thereby improving convergence stability and prediction accuracy. Second, 2nd-order Cartesian tensors explicitly represent anisotropic interatomic interactions and lattice deformations. Third, the layer-wise deep embedding of lattice information allows global information to be incorporated continuously into interatomic message passing, which appears important for capturing long-range effects.

Compared with existing relaxation frameworks, HotRelax offers several practical advantages. It predicts relaxed structures in a single forward pass, avoiding the iterative gradient-based optimization typically required by MLPs-based relaxation workflows and reducing inference time by orders of magnitude. At the same time, its architecture avoids complex components such as global lattice nodes, resulting in fewer parameters.

Overall, HotRelax addresses several limitations of existing end-to-end structure prediction models. By combining one-shot inference, prediction targets decoupling, high-order tensor representations, and lattice-aware message passing, it provides an efficient framework for relaxed-structure prediction across diverse materials systems. We expect it to be a useful tool for accelerating the discovery of next-generation functional materials across energy, catalysis, optoelectronics, and quantum technologies.

# Methods

## Implementation details

The HotRelax model is implemented in PyTorch. Benchmarks are performed on an NVIDIA RTX 5090 GPU with 32 GB of memory. The training objective is to minimize the sum of coordinate MAE $\mathcal{L}_{\text{coord}}$ and lattice MAE $\mathcal{L}_{\text{lattice}}$ between the HotRelax-predicted and DFT-relaxed structures, defined as:

$$\mathcal{L} = \mathcal{L}_{\text{coord}} + \mathcal{L}_{\text{lattice}} \tag{8}$$

$$\mathcal{L}_{\text{coord}} = \frac{1}{M}\frac{1}{N}\sum_{i=1}^{N}|\hat{\boldsymbol{r}}_i - \vec{\boldsymbol{r}}_i| \tag{9}$$

$$\mathcal{L}_{\text{lattice}} = \frac{1}{M}\frac{1}{9}\sum|\hat{\boldsymbol{L}} - \boldsymbol{L}| \tag{10}$$

where $M$ denotes the number of samples and $N$ denotes the number of atoms in a given sample, $\hat{\boldsymbol{r}}_i$ and $\vec{\boldsymbol{r}}_i$ are the predicted and reference Cartesian coordinates, $\hat{\boldsymbol{L}}$ and $\boldsymbol{L}$ are the predicted and reference lattice matrices.

For the IS2RE task on OC20, we implement the PaiNN[20] and Equiformer V2[23] models using the source code available at https://github.com/facebookresearch/fairchem/tree/c9971f72b793436e63685459cee0bf702ccf97c0. The associated experiments are conducted on four NVIDIA RTX 4090 GPUs, each with 24 GB of memory.

## DFT calculations

All DFT calculations were performed with VASP[43–46] using the generalized gradient approximation (GGA) with the Perdew-Burke-Ernzerhof (PBE) exchange-correlation functional[50]. The projector-augmented wave (PAW) method was adopted[51]. All calculations employed the electronic minimization algorithm with ALGO = All ("all band simultaneous update of orbitals"), a cutoff energy of 550 eV, an energy convergence criterion of $1.0\times10^{-5}$ eV, and a Gaussian smearing width of 0.02 eV. For the X-Mn-O dataset, self-consistent field (SCF) calculations were performed to obtain the total energy without spin polarization, using a 9×9×9 k-point mesh to ensure accurate total energies. The effective on-site Coulomb interaction parameter (U) for the Mn 3d orbitals was set to 3.9 eV, consistent with the value adopted in DeepRelax[27]. For the MP dataset, SCF calculations were performed with a 5×5×5 k-point mesh for structures containing fewer than 60 atoms and a 3×3×3 mesh for structures containing more than 60 atoms. All other parameters were chosen to be consistent with those of the MP by using the MPRelaxSet function in Pymatgen[42].

# Code Availability

The code for training and using the HotRelax model is open source, released under the MIT License. The code repository is accessible online, at: https://github.com/ShbYu/HotRelax.

# Data availability

The dataset for X-Mn-O is available at https://zenodo.org/records/8081655. The dataset for MP is available at https://figshare.com/articles/dataset/MPF_2021_2_8/19470599. The dataset for C2DB is available at https://2dhub.org/c2db/c2db.html. The dataset for 2DMD is available at https://zenodo.org/records/14027373. The dataset for OC20 is available at https://fair-chem.github.io/oc20.

# Acknowledgments

We are grateful for insightful discussions with Prof. Zachary Ulissi on the IS2RE task in the OC20 dataset. We would also like to thank Prof. Kristian Sommer Thygesen and Jens Jørgen Mortensen for generously providing the C2DB database, which includes both initial and DFT-relaxed structures. We gratefully acknowledge the financial support from the National Key R&D Program of China (grant no. 2022YFA1403201), the National Natural Science Foundation of China (grant number T2495231, 12125404, 123B2049), the Basic Research Program of Jiangsu (Grant BK20233001, BK20241253, BK20253009), the Jiangsu Funding Program for Excellent Postdoctoral Talent (2024ZB002, 2024ZB075), the Postdoctoral Fellowship Program of CPSF (Grant GZC20240695), the AI & AI for Science program of Nanjing University, and the Fundamental Research Funds for the Central Universities. The calculations were carried out using supercomputers at the High Performance

Computing Center of Collaborative Innovation Center of Advanced Microstructures, the high-performance supercomputing center of Nanjing University.

# Figures and tables

## Table 1 | The comparison results of PaiNN without DCL strategy, the PaiNN model with DCL strategy, and DeepRelax on the X-Mn-O dataset.

**Table 1 | The comparison results of PaiNN without DCL strategy, the PaiNN model with DCL strategy (PaiNN-DCL), and DeepRelax on the X-Mn-O dataset.** The evaluation metrics are based on mean absolute error of coordinates (MAE Coords.), lattice (MAE Latt.), and match rate (MR) between the predicted structure and the DFT-relaxed structure

| Model | MAE Coords. [Å]↓ | MAE Latt. [Å]↓ | MR [%]↑ |
|---|---|---|---|
| PaiNN[20] | 0.261 | 0.067 | 69.7 |
| PaiNN[20]-DCL | **0.114** | 0.066 | 84.2 |
| DeepRelax[27] | 0.116 | **0.063** | **84.6** |

# Fig. 1 | The architecture of HotRelax.

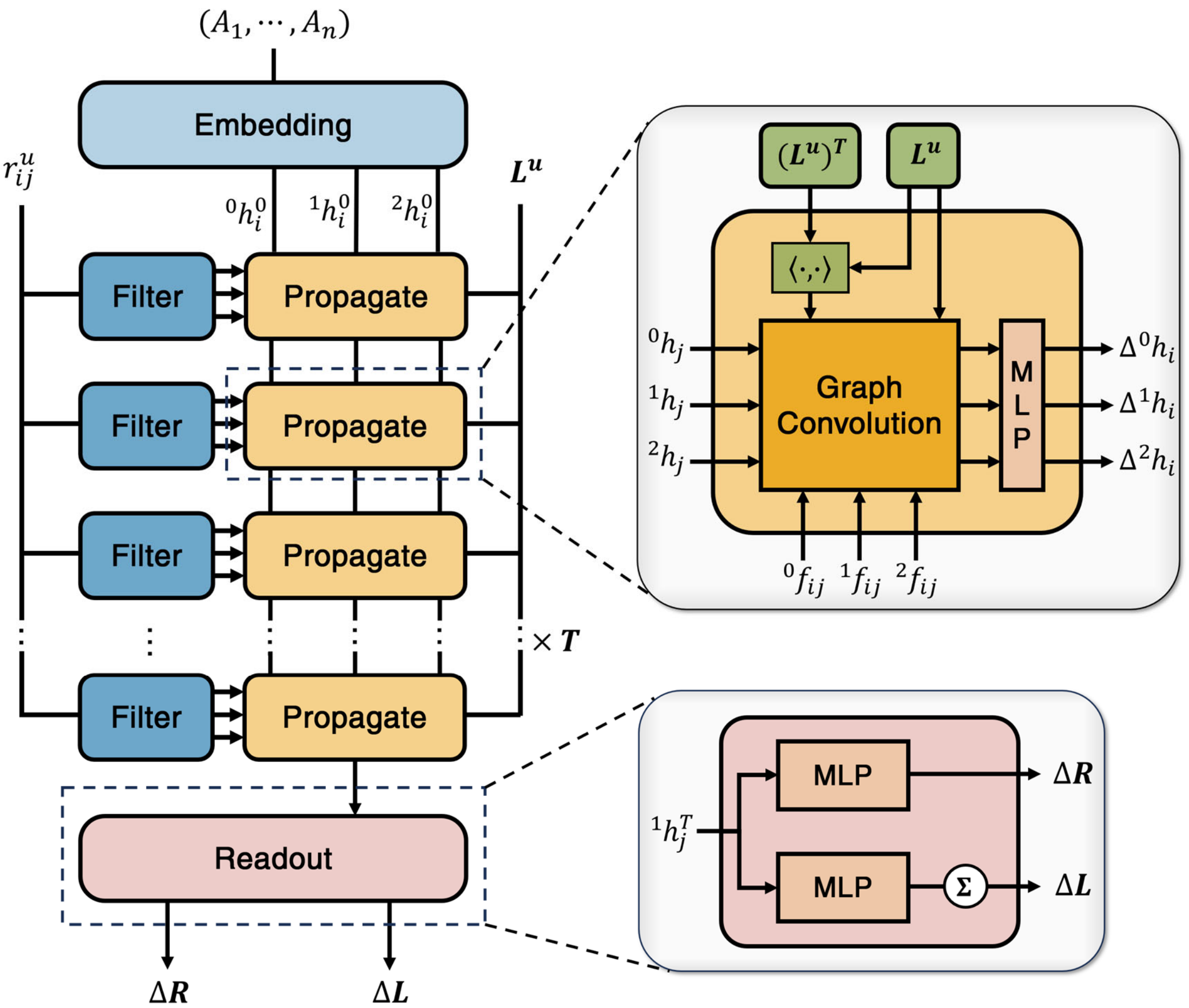

**Fig. 1 | The architecture of HotRelax.** HotRelax is derived from HotPP[24]. In the embedding layer, atomic numbers are embedded as initial scalar node features (${}^0h^0$), while the initial vector (${}^1h^0$) and second-order tensor (${}^2h^0$) node features are set to zero. The resulting node features, alongside edge features generated by the Filter layer, are fed into a sequence of T Propagate layers. Within each Propagate layer, a graph convolution operator acts on the node and edge features to update node states, by aggregating messages from neighboring nodes and edges with lattice vectors explicitly integrated into the convolution operation. The intermediate convolution outputs are subsequently passed through a multilayer perceptron (MLP), which produces incremental updates $\Delta{}^0h_i$, $\Delta{}^1h_i$ and $\Delta{}^2h_i$ for the scalar, vector, and tensor node features, respectively. After T full propagation blocks, the final output vector features are processed by an MLP and summed to generate the atomic coordinate updates ΔR and lattice vector updates ΔL.

# Table 2 | Comparison of HotRelax with baseline models on the X-Mn-O and MP datasets.

**Table 2 | Comparison of HotRelax with baseline models on the X-Mn-O and MP datasets.** Metrics are the MAE of Cartesian coordinates, lattice vectors and match rate between predicted and DFT-relaxed structures. Distributions of MAE are detailed in Supplementary Note 3. The best performance in each metric is highlighted in bold

| | X-Mn-O | | | MP | | |
|---|---|---|---|---|---|---|
| **Model** | **MAE Coords. [Å]↓** | **MAE Latt. [Å]↓** | **MR [%]↑** | **MAE Coords. [Å]↓** | **MAE Latt. [Å]↓** | **MR [%]↑** |
| Dummy | 0.314 | 0.221 | 64.8 | 0.096 | 0.072 | 95.6 |
| DeepRelax[27] | 0.116 | 0.063 | 84.3 | 0.066 | 0.042 | 95.6 |
| $E^3$Relax[30] | 0.105 | 0.066 | 85.6 | 0.058 | 0.042 | 96.0 |
| HotRelax | **0.097** | **0.058** | **86.7** | **0.055** | **0.040** | **96.6** |

# Table 3 | Comparison of HotRelax with baseline models on the C2DB dataset.

**Table 3 | Comparison of HotRelax with baseline models on the C2DB dataset.** Metrics are the MAE of Cartesian coordinates, lattice vectors and match rate between predicted and DFT-relaxed structures. Distributions of MAE are detailed in Supplementary Note 3. The best performance in each metric is highlighted in bold

| Model | MAE Coords. [Å]↓ | MAE Latt. [Å]↓ | MR [%]↑ |
|---|---|---|---|
| Dummy | 0.255 | 0.141 | 72.3 |
| DeepRelax[27] | 0.190 | 0.084 | 80.9 |
| $E^3$Relax[30] | 0.159 | 0.084 | 82.3 |
| HotRelax | **0.154** | **0.082** | **84.3** |

# Fig. 2 | Comparison of HotRelax with baseline models on the high-density defect structures.

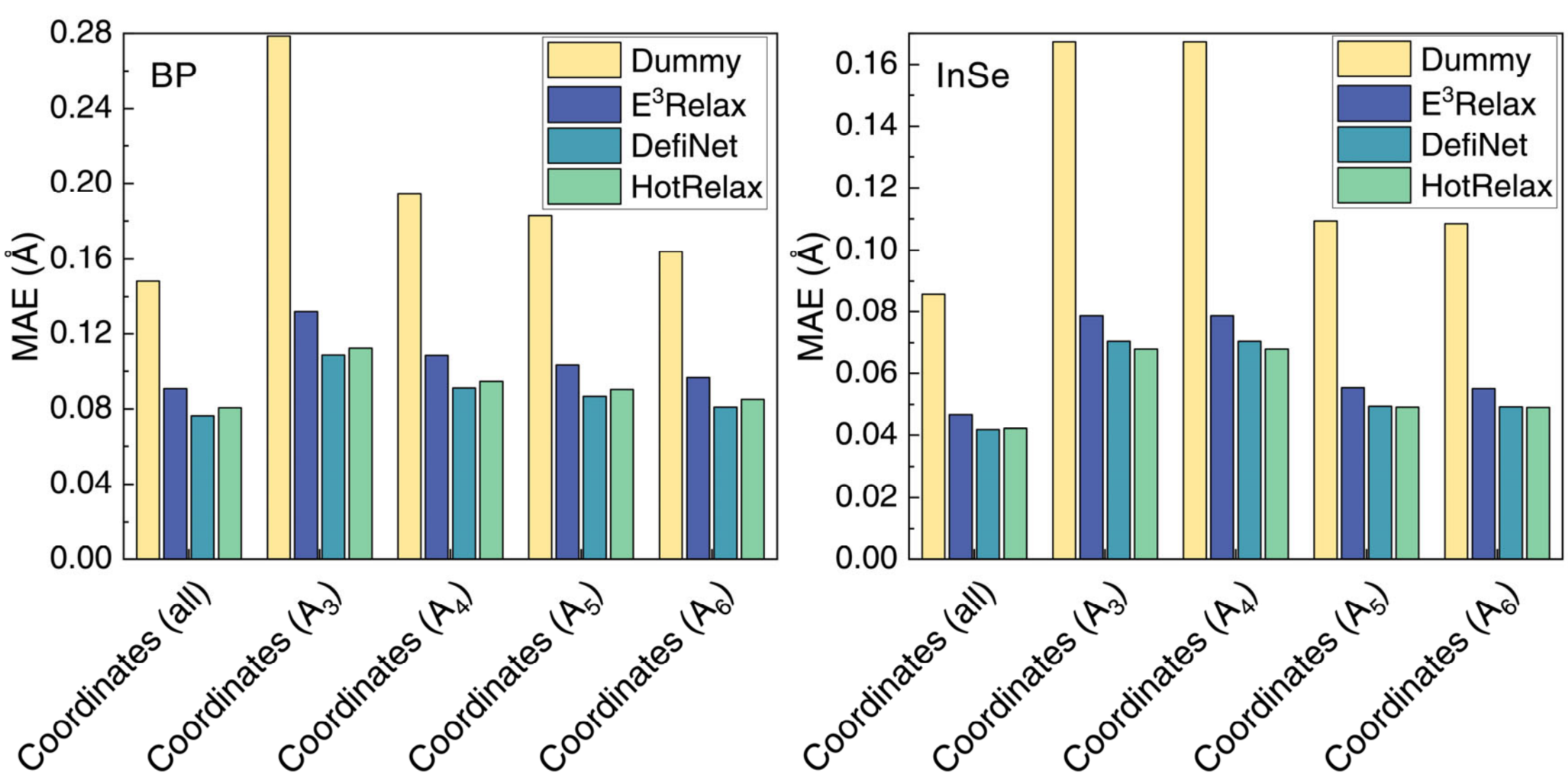

**Fig. 2 | Comparison of HotRelax with baseline models on the high-density defect structures.** Results of black phosphorus (BP) and InSe are shown here. Metrics are the MAE of Cartesian coordinates between predicted and DFT-relaxed structures. $A_3$, $A_4$, $A_5$, and $A_6$ denote the local MAE values calculated using only atoms within radii of 3, 4, 5, and 6 Å around the defect sites, respectively.

# Fig. 3 | Distribution of energy above relaxed state for HotRelax-predicted and unrelaxed structures.

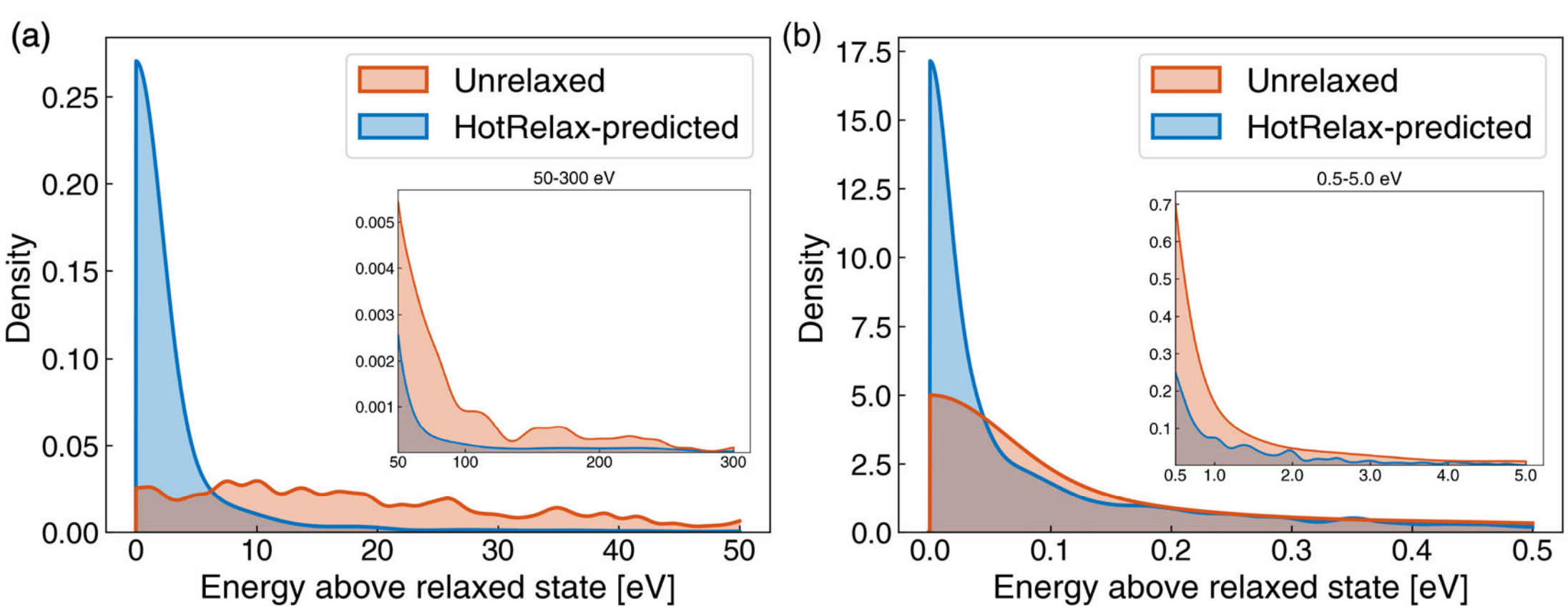

**Fig. 3 | Distribution of energy above relaxed state for HotRelax-predicted and unrelaxed structures. a** X-Mn-O and (**b**) MP dataset. Insets show magnified views of the high-energy tails (50–300 eV in (**a**) and 0.5–5.0 eV in (**b**)). Density curves were obtained using boundary-corrected reflection kernel density estimation to account for the non-negative support of the energy-above-relaxed-state values.

# Table 4 | Comparison of HotRelax with baseline models across four validation subsets on the OC20 dataset.

**Table 4 | Comparison of HotRelax with baseline models across four validation subsets on the OC20 dataset.** Metrics are the AMAE of Cartesian coordinates and match rate between predicted and DFT-relaxed structures. Distributions of AMAE are detailed in Supplementary Note 6. The best performance in each metric is highlighted in bold

| | AMAE Coords. [Å]↓ | | | | MR [%]↑ | | | |
|---|---|---|---|---|---|---|---|---|
| **Model** | **ID** | **OOD Ads** | **OOD Cat** | **OOD Both** | **ID** | **OOD Ads** | **OOD Cat** | **OOD Both** |
| Dummy | 0.245 | 0.240 | 0.247 | 0.201 | 24.4 | 28.4 | 22.9 | 32.2 |
| $E^3$Relax[30] | 0.147 | 0.152 | 0.150 | 0.125 | 52.7 | 49.2 | 52.3 | 57.7 |
| HotRelax | **0.139** | **0.145** | **0.141** | **0.117** | **55.8** | **52.4** | **55.4** | **60.8** |

## Fig. 4 | Comparison of structure relaxation with MLPs and HotRelax.

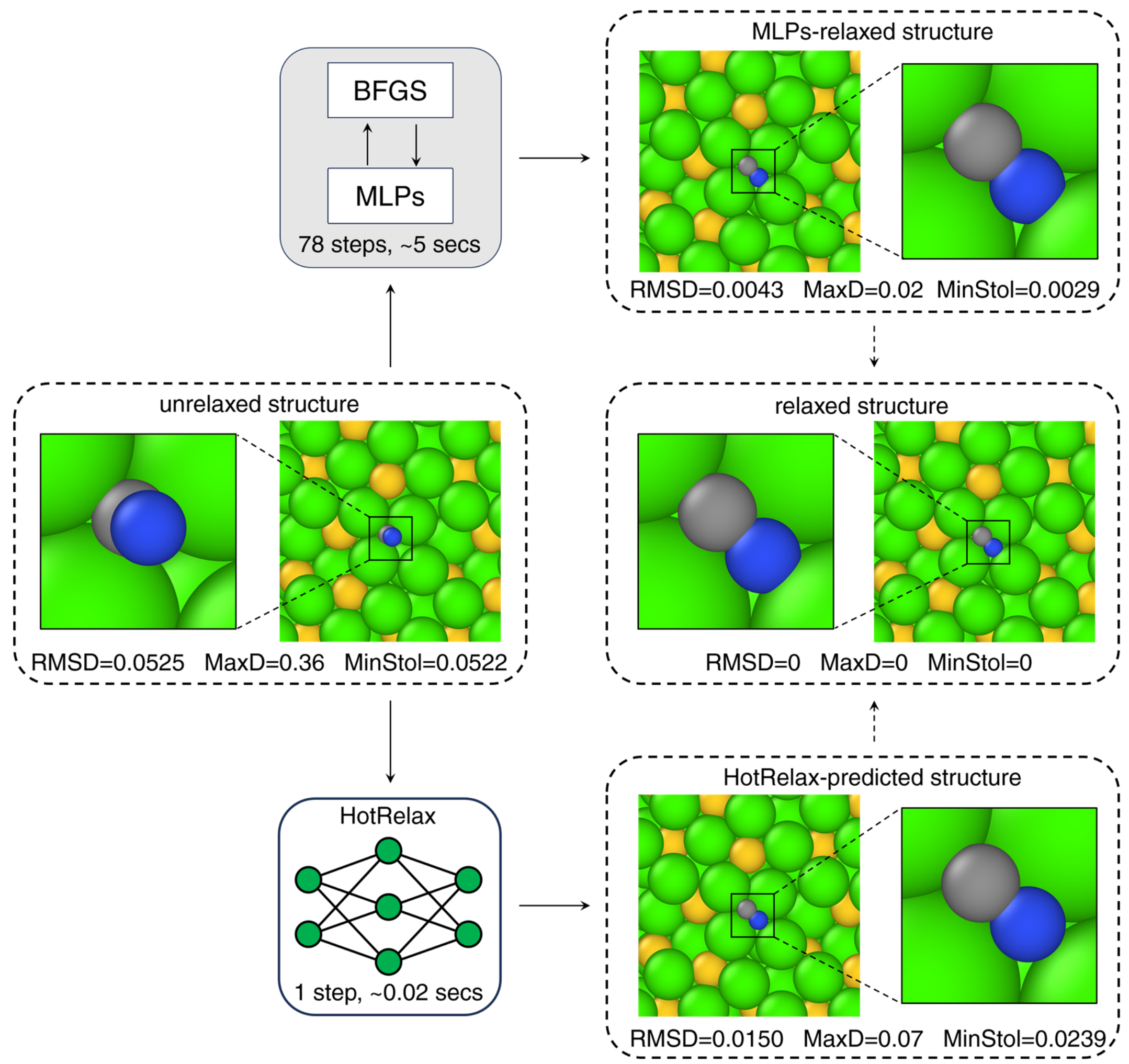

**Fig. 4 | Comparison of structure relaxation with MLPs and HotRelax.** The unrelaxed structure differs markedly from the relaxed reference. MLPs require 78 steps (~5 s) to reach a matched configuration, while HotRelax achieves a matched structure in 1 step (~0.02 s). Metrics are normalized RMSD between predicted and relaxed structures, maximum displacement (MaxD) of paired atomic sites, and minimum valid match tolerance (MinStol). All values were calculated via Pymatgen[42].

## Fig. 5 | IS2RE workflows.

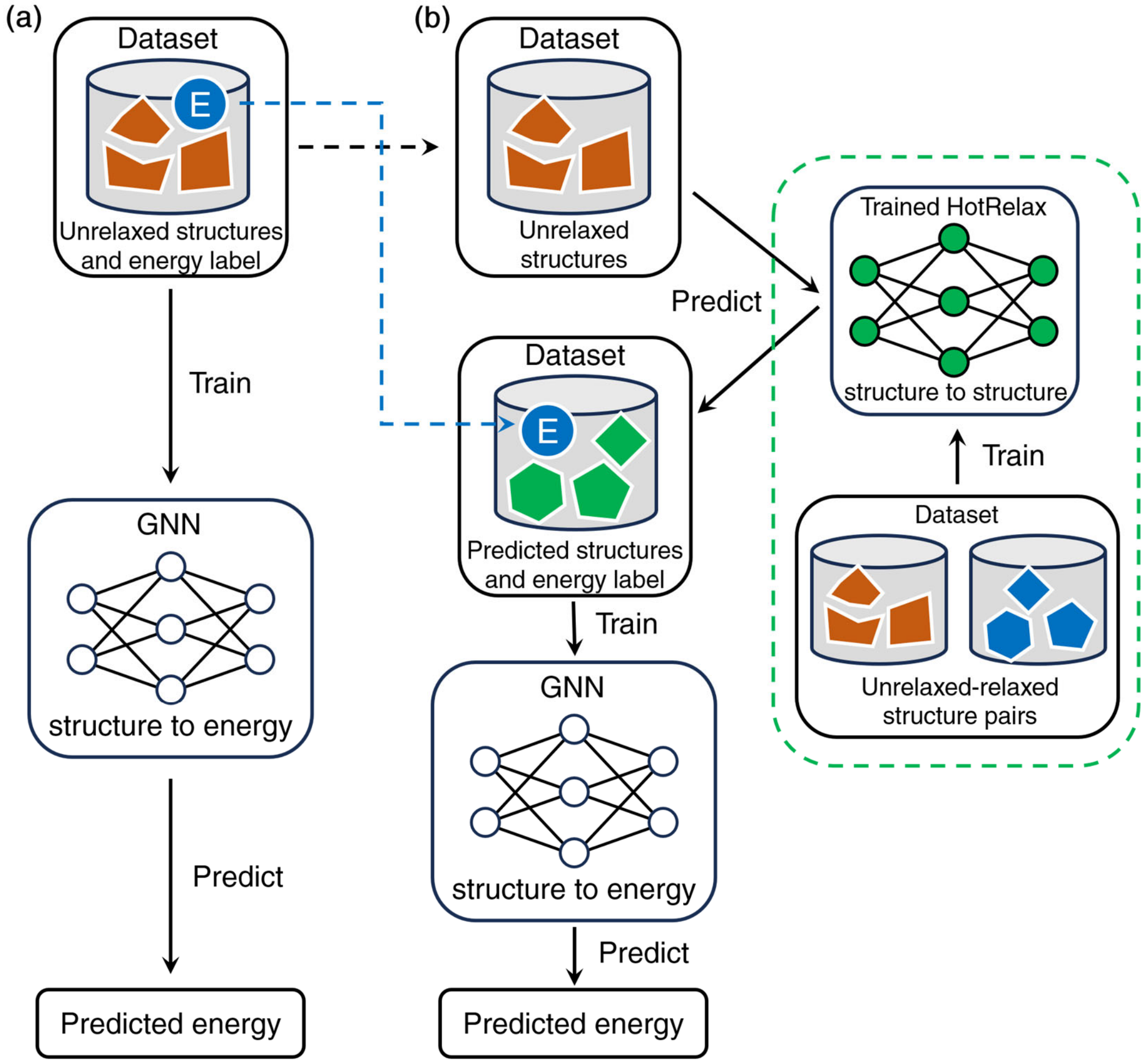

**Fig. 5 | IS2RE workflows. a** Standard IS2RE workflow. The energy of the DFT-relaxed structure is directly predicted from unrelaxed input structures via a GNN. **b** HotRelax-based two-stage IS2RE workflow. Stage 1: HotRelax is trained on unrelaxed-relaxed structure pairs from the OC20 training set. Stage 2: pre-trained HotRelax generates predicted structures from all unrelaxed inputs, which are paired with the energies of the corresponding DFT-relaxed structures to train the energy prediction model. For inference, input structures are first processed by HotRelax to yield predicted configurations, which are then fed into the energy prediction model for relaxed-state energy prediction.

# Table 5 | Comparison of different IS2RE workflows.

**Table 5 | Comparison of different IS2RE workflows.** Metric is the MAE of energies. Distributions are detailed in Supplementary Note 8. Models are evaluated across four validations and the average values across four test subsets

| | Energy MAE [eV]↓ | | | | |
|---|---|---|---|---|---|
| **Model** | **ID** | **OOD Ads** | **OOD Cat** | **OOD Both** | **Test Ave.** |
| PaiNN[20] | 0.575 | 0.655 | 0.573 | 0.596 | 0.656 |
| HotRelax-PaiNN | **0.519** | **0.558** | **0.506** | **0.491** | **0.552** |
| EquiformerV2[23] | 0.538 | 0.668 | 0.539 | 0.598 | 0.626 |
| HotRelax-EquiformerV2 | **0.478** | **0.552** | **0.478** | **0.480** | **0.547** |